\documentclass[aps,pra,floatfix,superscriptaddress,showpacs,twocolumn]{revtex4}
\usepackage{amsmath}
\usepackage{graphicx}

\begin{document}

\title{Production and detection of atomic hexadecapole at Earth's magnetic field}

\author{V.~M.~Acosta}
\affiliation{Department of Physics, University of California,
Berkeley, CA 94720-7300}
\author{M.~Auzinsh}
\affiliation{Department of Physics, University of Latvia, 19 Rainis
blvd, Riga, LV-1586, Latvia}
\author{W.~Gawlik}
\affiliation{Center for Magneto-Optical Research, Institute of
Physics, Jagiellonian University, Reymonta 4, 30-059 Krak\'ow,
Poland}
\author{P.~Grisins}
\affiliation{Department of Physics, University of Latvia, 19 Rainis
blvd, Riga, LV-1586, Latvia}
\author{J.~M.~Higbie}
\affiliation{Department of Physics, University of California,
Berkeley, CA 94720-7300}
\author{D.~F.~Jackson~Kimball}
\affiliation{Department of Physics, California State University --
East Bay, 25800 Carlos Bee Blvd., Hayward, CA 94542, USA}
\author{L.~Krzemien}
\affiliation{Center for Magneto-Optical Research, Institute of
Physics, Jagiellonian University, Reymonta 4, 30-059 Krak\'ow,
Poland}
\author{M.~P.~Ledbetter}
\affiliation{Department of Physics, University of California,
Berkeley, CA 94720-7300}
\author{S.~Pustelny}
\affiliation{Center for Magneto-Optical Research, Institute of
Physics, Jagiellonian University, Reymonta 4, 30-059 Krak\'ow,
Poland}
\author{S.~M.~Rochester}
\affiliation{Department of Physics, University of California,
Berkeley, CA 94720-7300}
\author{V.~V.~Yashchuk}
\affiliation{Advanced Light Source Division, Lawrence Berkeley
National Laboratory, Berkeley CA 94720, USA}
\author{D.~Budker}
\email{budker@berkeley.edu} \affiliation{Department of Physics,
University of California, Berkeley, CA 94720-7300}
\affiliation{Nuclear Science Division, Lawrence Berkeley National
Laboratory, Berkeley CA 94720, USA}

\maketitle

\textbf{Anisotropy of atomic states is characterized by population
differences and coherences between Zeeman sublevels. It can be
efficiently created and probed via resonant interactions with light,
the technique which is at the heart of modern atomic clocks
\cite{VanierAudoin} and magnetometers \cite{BudRom2007}.
Polarization moments (PM) characterizing anisotropy of a state with
total angular momentum $F$ are coefficients in the expansion of the
density matrix into irreducible tensor operators of rank
$\kappa=0,...,2F$ and projection $q=-\kappa,\ldots\kappa$. The
lowest PMs are population ($\kappa=0$), orientation ($\kappa=1$),
and alignment ($\kappa=2$). Recently, nonlinear magneto-optical
techniques have been developed for selective production and
detection of higher PMs, hexadecapole ($\kappa=4$) and
hexacontatetrapole ($\kappa=6$), in the ground states of the alkali
atoms \cite{Yas2003Select,Pus2006pp}. Extension of these techniques
into the range of geomagnetic fields is important for practical
applications. This is because hexadecapole polarization
corresponding to the $\Delta M=4$ Zeeman coherence, with maximum
possible $\Delta M$ for electronic angular momentum $J=1/2$ and
nuclear spin $I=3/2$, is insensitive to the nonlinear Zeeman effect
(NLZ). This is of particular interest because NLZ normally leads to
resonance splitting and systematic errors in atomic magnetometers.
However, optical signals due to the hexadecapole moment decline
sharply as a function of magnetic field.
In this Letter, we report a novel method that allows selective
creation of a macroscopic long-lived ground-state hexadecapole
polarization.  The immunity of the hexadecapole signal to NLZ is
demonstrated with $F=2$ $^{87}$Rb atoms at Earth's field.}

\begin{figure}
\includegraphics[scale=.5]{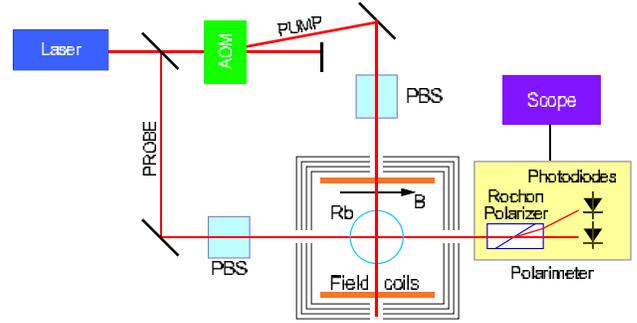}
    \caption{Schematic of the experimental apparatus. The modulated pump light beam is used to polarize $^{87}$Rb
    atoms. The unmodulated probe beam is used to detect the evolution
    of the atomic polarization moments via induced optical rotation. AOM: acousto-optic modulator; PBS: polarizing beam
    splitters. See Methods.}
    \label{Fig_apparatus}
\end{figure}

Figure \ref{Fig_apparatus} shows a schematic of the experimental
apparatus. A diode laser tuned to the D1 line interacts with
$^{87}$Rb atoms contained in an antirelaxation coated cell, to which
a magnetic field is applied. The $F=2$ ground-state atoms are pumped
using a sequence of short pulses of linearly polarized light with a
repetition rate determined by the Larmor precession frequency of the
atomic angular momentum. At the end of the sequence, the pump light
is blocked. The evolution of the atomic polarization is observed by
measuring the angle of optical rotation of an unmodulated probe beam
whose initial polarization is the same as that of the pump. An
alternative arrangement where the pump beam itself was attenuated
after the pumping cycle and used for probing (without modulation)
was also used in some of the experiments.

\begin{figure}
    \includegraphics[width=3 in]{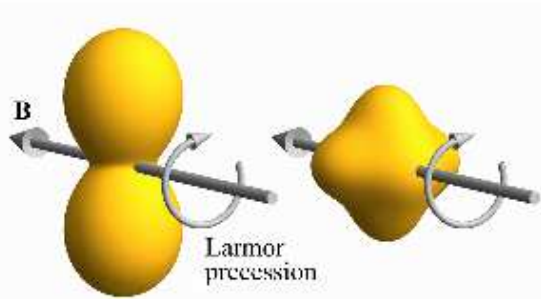}
    \caption{Angular momentum probability surfaces for quadrupole (left) and hexadecapole (right) for $F=2$.
    The shown polarizations are
    transverse to the quantization axis, which is chosen along the magnetic field.
    The anisotropies and polarization surfaces undergo
    Larmor precession around the magnetic field.}
    \label{Fig_pea}
\end{figure}

The symmetry properties of polarization moments can be illustrated
using angular-momentum probability surfaces
\cite{Auz97,Roc2001,Ale2005}, in which the distance between the
origin and the surface in a given direction is proportional to the
probability of finding maximal projection $m=F$ in that direction.
Figure \ref{Fig_pea} shows such surfaces for the quadrupole
($\kappa=2$) and hexadecapole ($\kappa=4$) moments. The surfaces
shown correspond to combinations of $q=-2$ and 2 for the quadrupole,
and $q=-4$ and 4 for the hexadecapole with the quantization axis
along the magnetic field \footnote{In terms of coherences between
Zeeman sublevels, these anisotropies correspond to $\Delta M=2$ and
$\Delta M=4$ coherences, respectively.}. The atomic angular momentum
precesses at the Larmor frequency $\Omega_L$ proportional to the
applied magnetic field. Consider the ``peanut''-shaped quadrupole
moment. When the peanut has rotated by an angle $\pi$, it is
impossible to differentiate it from its initial state, i.e., it has
a 2-fold symmetry. Consequently, efficient pumping of the quadrupole
is achieved with light modulated at an angular frequency
$2\Omega_L$. Precession of the quadrupole results in optical
rotation of the probe light oscillating at $2\Omega_L$. In general,
a PM with a component $q$ has $|q|$-fold symmetry and can therefore
be pumped with light harmonically modulated at $|q|\Omega_L$ or with
short pulses at a repetition rate of $|q|\Omega_L/(2\pi n)$, where
$n$ is an integer. The polarization moments prepared in this way
produce an optical rotation signal at $|q|\Omega_L$
\cite{Yas2003Select,Pus2006pp}. Polarization moments with $|q|$-fold
symmetry can be detected by demodulating the optical rotation signal
at $|q|\Omega_L$.

\begin{figure}
    \includegraphics[scale=0.5]{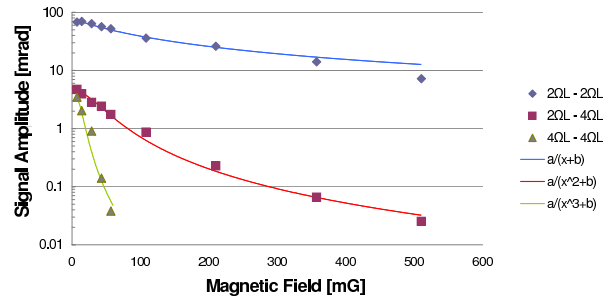}
    \caption{Magnetic-field dependence of optical rotation amplitudes for
    quadrupole (blue diamonds),
    hexadecapole pumped with light modulated at $2\Omega_L$ (maroon squares),
    and hexadecapole pumped at $4\Omega_L$ (green triangles).
    The quadrupole and hexadecapole, pumped at $2\Omega_L$ decrease much more slowly with magnetic
    field compared to hexadecapole pumped at $4\Omega_L$. The solid lines are fits by ad-hoc
    functions.}
    \label{Fig_magfield}
\end{figure}

\begin{figure}
    \includegraphics[width=9 cm]{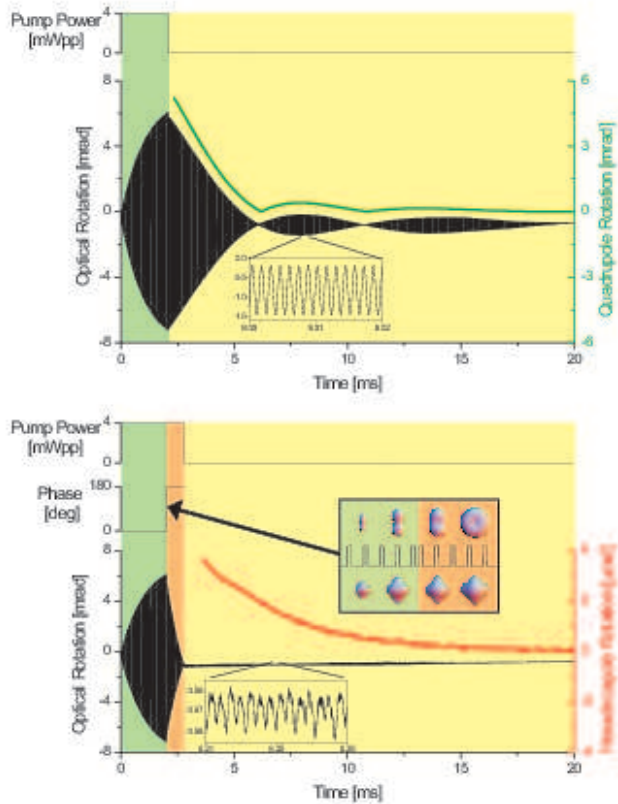}
    \caption{Comparison of the demodulated quadrupole signal (top) without phase flip (see text) and of the
    hexadecapole signal
    obtained with phase flip (bottom).
    For the case of pumping without phase flip (top), the signal demodulated at $2
    \Omega_{L}$ is plotted alongside the raw optical
    rotation signal. The inset on the top plot is a blowup of the raw
    optical-rotation signal during a revival stage of the
    quadrupole-signal beats
    occurring due to the nonlinear Zeeman effect (NLZ). This signal on the bottom plot (with phase flip) is
    demodulated at $4\Omega_{L}$, and the resulting curve shows the
    absence of the beats related to NLZ. The inset on the right
    shows the details of the pumping pulses near the phase flip, along with the
    angular-momentum probability surfaces characterizing the ensemble at different stages of pumping. The magnetic field around which these
    surfaces precess is normal to the page.}
    \label{Fig_bigplot}
\end{figure}

By pumping at $4\Omega_L$, hexadecapole can be created without the
presence of the quadrupole \cite{Yas2003Select,Pus2006pp}. This is
desirable, as the presence of the typically much larger quadrupole
signal makes it difficult to observe the hexadecapole.
Unfortunately, the amount of hexadecapole pumped in this way rapidly
decreases with the magnetic field strength, as seen in Figure
\ref{Fig_magfield} (bottom trace). This can be understood by
recalling that a photon is a spin-one particle, so it is described
by tensor operators of rank $\kappa\leq 2$. Consequently, two
photons are needed to produce an atomic state of $\kappa=4$
(hexadecapole). When pumping at $4\Omega_L$, both photons must
interact with the atoms in a single pulse in order to create
hexadecapole. The alternative mode of hexadecapole creation, two
single-photon processes occurring during separate cycles, is
ineffective when pumping in this manner, as it would require the
survival of quadrupole between the pumping pulses. In fact, each
successive pumping cycle creates a quadrupole that is orthogonal to
the quadrupole that was created in the previous cycle. This is
because the latter has rotated by $\pi/2$ due to Larmor precession
over the duration of the cycle. Thus, the quadrupole polarizations
from successive cycles cancel the transverse quadrupole (and the net
result is a longitudinal quadrupole, seen as a ``doughnut'' in the
inset of Fig. \ref{Fig_bigplot}, which does not aid in creation of
transverse hexadecapole). As the field increases, the pumping period
[$T=2\pi/(4\Omega_L)$] decreases, reducing the probability of a
process occurring which involves interactions with two photons in a
single pulse.
Even using the strongest pump power available in the experiment ($4\
$mW), the hexadecapole signals cannot be distinguished above noise
at the Earth's field ($4\Omega_L/2\pi \approx 1.4\ $MHz) in this
scheme.

An alternative method for efficiently pumping hexadecapole is to use
light modulated at $2\Omega_L$, producing both quadrupole and
hexadecapole. In this scheme, the requirement of pumping the
hexadecapole in a single pulse is alleviated as the hexadecapole can
be obtained by ``promoting" the quadrupole polarization with just a
single-photon interaction. This method allows one to obtain
hexadecapole signals that, while they still decrease with magnetic
field in the present experiment,
nevertheless remain observable at the Earth field (Fig.\
\ref{Fig_magfield}). The remaining decrease in the hexadecapole
signal is related to nonlinearity of the interaction, in this case,
at the probing stage. The decrease is qualitatively reproduced by a
model density-matrix calculation to be described elsewhere. Note
that because we use an antirelaxation-coated cell, atoms pumped into
a state with quadrupole polarization may leave the light beam and
bounce around the cell for a relatively long time (on the order of
the polarization-relaxation time), before they interact with the
light again and are pumped into a state with hexadecapole
polarization. Note also that the decrease of the quadrupole signal
with magnetic field seen in Fig.\ \ref{Fig_magfield} is a
consequence of nonlinearity in the probing process at the relatively
high powers required for probing hexadecapole (and which we use for
quadrupole to be able to compare the two signals directly). With the
single-beam arrangement, we verified that the quadrupole signal did
not drop for magnetic fields up to 268 mG with a low probe-light
power of 3\ $\mu$W.

Figure \ref{Fig_bigplot} (top) shows the optical rotation signal
obtained with 1500 square pump pulses with repetition rate of
$2\Omega_L/(2\pi)\approx 714\ $kHz and a duration of 1/8 of a
period. Significant beating of the signal is observed after the
pulse sequence ends, which is due to NLZ (the effect of NLZ on
nonlinear magneto-optical rotation is discussed in Ref.
\cite{Aco2006}). Fits of the demodulated quadrupole signal indicate
the presence of three close frequencies in the time dependence of
the optical rotation with a splitting between adjacent frequencies
of $\approx 72\ $Hz close to the value of the NLZ splitting,
$\delta_{NLZ}=74.65\ $Hz for this field. The overall exponential
decay time ($\tau\approx 4.7\ $ms) is determined by the relaxation
of the PMs due to dephasing from collisions and magnetic-field
inhomogeneities, and by residual probe-power broadening. Buried
under the much larger quadrupole signal there is also hexadecapole
signal producing modulation of the optical rotation at $4\Omega_L$.
Unfortunately, due to intrinsic nonlinearities in the detection
electronics as well as those due to the interaction of atoms with a
strong probe light \cite{Ser66}, the large quadrupole signal leads
to the presence of a ``false hexadecapole'' signal at $4\Omega_L$,
and it is hard to distinguish the two contributions.

The solution implemented in the present work is to eliminate the
quadrupole just before probing. To accomplish this, the pumping is
separated into two stages. The first stage is the same as in the
pumping scheme above: the atoms are pumped at $2\Omega_L$ for
$\approx 1500$ cycles. Then, the phase of the pumping is flipped by
$\pi$ (see inset in the bottom plot of Fig. \ref{Fig_bigplot}). The
quadrupole is now pumped orthogonally to its previous alignment and
the resulting sum of orthogonal quadrupole moments leaves no net
transverse alignment (the resulting doughnut shape, corresponding to
longitudinal alignment that causes no optical rotation of the probe
light, is shown inset at the bottom plot in Fig. \ref{Fig_bigplot}).
However, the hexadecapole produced before and after the phase flip
is identical, so it continues to be pumped even after the phase
flip. After $\approx 500$ cycles, the quadrupole reaches a minimum
and the pump is shut off. Figure \ref{Fig_bigplot} (bottom)
demonstrates the phase-flip pumping scheme and the resulting signal.
The quadrupole signal (not shown) is reduced by about a factor of 40
in this scheme, allowing for the reliable recovery of the signal due
to the hexadecapole moment. The demodulated hexadecapole signal is a
simple exponential decay ($\tau=4.2$ ms), clearly demonstrating the
absence of NLZ-induced beating.

\begin{figure}
    \includegraphics[width=7.5 cm]{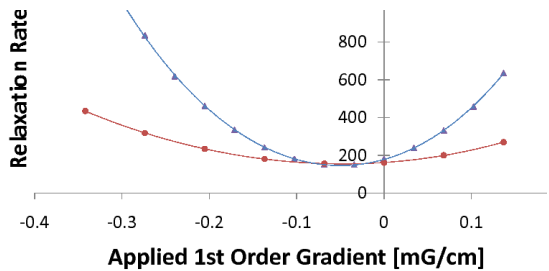}
    \caption{The dependence of the observed relaxation rates for the quadrupole and hexadecapole signals
    on the magnetic-field gradient applied in the direction of the magnetic field. In both cases,
    pumping was with light pulsed at a repetition rate of $2\Omega_L/(2\pi)$ at $B\approx 108\ $mG,
    and the relaxation was determined from the optical rotation after the pump light was shut off.
    Solid lines show fits to fourth-order polynomials; the ratio of curvatures near the vertex is consistent with
    the expected four times higher sensitivity of the hexadecapole to gradients \cite{Pus2006grad}.}
    \label{Fig_gradients}
\end{figure}

In an additional series of measurements, we studied the dependence
of the quadrupole and hexadecapole optical-rotation signals on the
gradient of the magnetic field. These measurements (Fig.
\ref{Fig_gradients}) confirmed the expected (see, for example, Ref.
\cite{Pus2006grad}) four-times higher sensitivity of the
hexadecapole to field gradients compared to the quadrupole.

In this work we have demonstrated a way to create and detect
macroscopic hexadecapole polarization (corresponding to a $\Delta
M=4$ Zeeman coherence) in the geomagnetic field range. The amount of
hexadecapole created at these fields was dramatically enhanced by
pumping at $2\Omega_L$, the frequency associated with efficient
production of the (lower-rank) quadrupole moment. Phase flipping
allowed the elimination of the large quadrupole signal in order to
effectively uncover the smaller hexadecapole signal. The resulting
hexadecapole signal demonstrates the absence of beating associated
with the nonlinear Zeeman effect. The NLZ-free hexadecapole signals
are attractive for applications in atomic magnetometry because the
linear relation between the magnetic field and the spin-precession
frequency is maintained over the geophysical field range. However, a
shortcoming of the present method from the point of view of
practical applications is the relative smallness of the hexadecapole
signal. This is related to the necessity of employing a nonlinear
interaction for probing hexadecapole polarization via optical
rotation (similar to the case of pumping) \cite{Yas2003Select}. In
future work, it may be possible to overcome this by monitoring the
hexadecapole moment via fluorescence detection
\cite{Duc73,Auz86,Oku2001}.

\section*{Acknowledgements}

The authors acknowledge stimulating discussions with T.\ Karaulanov,
and E.\ Corsini. This work has been supported by an ONR MURI
program, NSF, and KBN grant \# 1 P03B 102 30. S.P. is a scholar of
the Foundation for Polish Science.

\section*{Methods}
A New Focus Vortex diode laser tuned to the $^{87}$Rb $F=2
\rightarrow F'=2$ transition on the D1 (795 nm) line was separated
into two beams: a 4-mW linearly polarized pump beam modulated with
an acousto-optic modulator (AOM, driven at 80 MHz) and a continuous
probe beam with the same initial polarization. The probe power was
25 $\mu$W for the data of Figs. \ref{Fig_bigplot} and
\ref{Fig_magfield} and 9 $\mu$W for the data of Fig.
\ref{Fig_gradients}.  The laser frequency was fine-tuned to maximize
the hexadecapole signal. Additional data were taken with a single
laser beam serving as both pump and probe, tuned to the $F=2
\rightarrow F'=1$ transition.
The Rb atoms were housed in an evacuated glass cell with paraffin
antirelaxation coating \cite{AleLIAD} contained in an oven
maintaining the cell at $42^\circ$C. Four layers of $\mu$-metal
shielding and a multi-order gradient-coil system were used to
maintain a stable magnetic field of $510\ $mG. The angle of
polarization of the outgoing probe beam was measured using a Rochon
polarizing beam splitter and a high-speed, large area
photodiode-op-amp circuit.
\bibliography{NMObibl}

\begin{thebibliography}{10}
\expandafter\ifx\csname url\endcsname\relax
  \def\url#1{\texttt{#1}}\fi
\expandafter\ifx\csname urlprefix\endcsname\relax\def\urlprefix{URL }\fi
\providecommand{\bibinfo}[2]{#2}
\providecommand{\eprint}[2][]{\url{#2}}

\bibitem{VanierAudoin}
\bibinfo{author}{Vanier, J.} \& \bibinfo{author}{Audoin, C.}
\newblock \emph{\bibinfo{title}{The quantum physics of atomic frequency
  standards}} (\bibinfo{publisher}{A. Hilger}, \bibinfo{address}{Bristol ;
  Philadelphia}, \bibinfo{year}{1989}).

\bibitem{BudRom2007}
\bibinfo{author}{Budker, D.} \& \bibinfo{author}{Romalis, M.~V.}
\newblock \bibinfo{title}{Optical magnetometry}.
\newblock \emph{\bibinfo{journal}{Nature Physics}}
  \textbf{\bibinfo{volume}{3}}, \bibinfo{pages}{227--234}
  (\bibinfo{year}{2007}).

\bibitem{Yas2003Select}
\bibinfo{author}{Yashchuk, V.~V.} \emph{et~al.}
\newblock \bibinfo{title}{Selective addressing of high-rank atomic polarization
  moments}.
\newblock \emph{\bibinfo{journal}{Phys. Rev. Lett.}}
  \textbf{\bibinfo{volume}{90}}, \bibinfo{pages}{253001}
  (\bibinfo{year}{2003}).

\bibitem{Pus2006pp}
\bibinfo{author}{Pustelny, S.} \emph{et~al.}
\newblock \bibinfo{title}{Pump-probe nonlinear magneto-optical rotation with
  frequency-modulated light}.
\newblock \emph{\bibinfo{journal}{Phys. Rev. A}} \textbf{\bibinfo{volume}{73}},
  \bibinfo{pages}{023817} (\bibinfo{year}{2006}).

\bibitem{Auz97}
\bibinfo{author}{Auzinsh, M.}
\newblock \bibinfo{title}{Angular momenta dynamics in magnetic and electric
  field: classical and quantum approach}.
\newblock \emph{\bibinfo{journal}{Can. J. Phys.}}
  \textbf{\bibinfo{volume}{75}}, \bibinfo{pages}{853--72}
  (\bibinfo{year}{1997}).

\bibitem{Roc2001}
\bibinfo{author}{Rochester, S.~M.} \& \bibinfo{author}{Budker, D.}
\newblock \bibinfo{title}{Atomic polarization visualized}.
\newblock \emph{\bibinfo{journal}{Am. J. Phys.}} \textbf{\bibinfo{volume}{69}},
  \bibinfo{pages}{450--4} (\bibinfo{year}{2001}).

\bibitem{Ale2005}
\bibinfo{author}{Alexandrov, E.~B.} \emph{et~al.}
\newblock \bibinfo{title}{Dynamic effects in nonlinear magneto-optics of atoms
  and molecules: review}.
\newblock \emph{\bibinfo{journal}{J. Opt. Soc. Am. B}}
  \textbf{\bibinfo{volume}{22}}, \bibinfo{pages}{7--20} (\bibinfo{year}{2005}).

\bibitem{Aco2006}
\bibinfo{author}{Acosta, V.} \emph{et~al.}
\newblock \bibinfo{title}{Nonlinear magneto-optical rotation with
  frequency-modulated light in the geophysical field range}.
\newblock \emph{\bibinfo{journal}{Phys. Rev. A}} \textbf{\bibinfo{volume}{73}},
  \bibinfo{pages}{053404} (\bibinfo{year}{2006}).

\bibitem{Ser66}
\bibinfo{author}{Series, G.~W.}
\newblock \bibinfo{title}{Theory of the modulation of light in optical pumping
  experiments}.
\newblock \emph{\bibinfo{journal}{Proc. Phys. Soc.}}
  \textbf{\bibinfo{volume}{88}}, \bibinfo{pages}{957--968}
  (\bibinfo{year}{1966}).

\bibitem{Pus2006grad}
\bibinfo{author}{Pustelny, S.}, \bibinfo{author}{Jackson~Kimball, D.~F.},
  \bibinfo{author}{Rochester, S.~M.}, \bibinfo{author}{Yashchuk, V.~V.} \&
  \bibinfo{author}{Budker, D.}
\newblock \bibinfo{title}{Influence of magnetic-field inhomogeneity on
  nonlinear magneto-optical resonances}.
\newblock \emph{\bibinfo{journal}{Phys. Rev. A}} \textbf{\bibinfo{volume}{74}},
  \bibinfo{pages}{063406} (\bibinfo{year}{2006}).

\bibitem{Oku2001}
\bibinfo{author}{Okunevich, A.~I.}
\newblock \bibinfo{title}{On the possibility of detecting the transverse
  component of the hexadecapole moment of atoms in fluorescent emission}.
\newblock \emph{\bibinfo{journal}{Opt. Spectrosc.}}
  \textbf{\bibinfo{volume}{91}}, \bibinfo{pages}{177--83}
  (\bibinfo{year}{2001}).

\bibitem{Auz86}
\bibinfo{author}{Auzinsh, M.~P.}, \bibinfo{author}{Tamanis, M.~Y.} \&
  \bibinfo{author}{Ferber, R.~S.}
\newblock \bibinfo{title}{Zeeman quantum beats after optical depopulation of
  the ground electronic state of diatomic molecules}.
\newblock \emph{\bibinfo{journal}{Sov. Phys. JETP}}
  \textbf{\bibinfo{volume}{63}}, \bibinfo{pages}{688--693}
  (\bibinfo{year}{1986}).

\bibitem{Duc73}
\bibinfo{author}{Ducloy, M.}
\newblock \bibinfo{title}{Nonlinear effects in optical pumping of atoms by a
  high-intensity multimode gas laser. {G}eneral theory}.
\newblock \emph{\bibinfo{journal}{Phys. Rev. A}} \textbf{\bibinfo{volume}{8}},
  \bibinfo{pages}{1844--59} (\bibinfo{year}{1973}).

\bibitem{AleLIAD}
\bibinfo{author}{Alexandrov, E.~B.} \emph{et~al.}
\newblock \bibinfo{title}{Light-induced desorption of alkali-metal atoms from
  paraffin coating}.
\newblock \emph{\bibinfo{journal}{Phys. Rev. A}} \textbf{\bibinfo{volume}{66}},
  \bibinfo{pages}{042903/1--12 [Erratum: Phys. Rev. A \textbf{70}, 049902(E)
  (2004)]} (\bibinfo{year}{2002}).

\end{thebibliography}
\bibliographystyle{naturemag}

\end{document}